\documentclass[11pt,onecolumn]{IEEEtran}
\usepackage{graphicx}
\usepackage{subfigure}
\usepackage{epsfig}

\newcommand{\vp}{\mbox{${\bf p}$}}

\newcommand{\vx}{\mbox{${\bf x}$}}

\newcommand{\vn}{\mbox{${\bf n}$}}

\newcommand{\mB}{\hbox{{\bf B}}}

\newcommand{\mC}{\hbox{{\bf C}}}

\newcommand{\mW}{\hbox{{\bf W}}}

\newcommand{\ga}{\alpha}
\newcommand{\gb}{\beta}
\newcommand{\grg}{\gamma}
\newcommand{\gd}{\delta}
\newcommand{\gre}{\varepsilon}

\newcommand{\gl}{\lambda}

\newcommand{\gr}{\rho}

\newcommand{\gt}{\tau}

\newcommand{\SNR}{\ensuremath{\hbox{SNR}}}

\newtheorem{theorem}{Theorem}[section]
\newtheorem{lemma}[theorem]{Lemma}

\newtheorem{prop}{Proposition}[section]
\newtheorem{claim}{Claim}[section]

\newtheorem{definition}{Definition}[section]
\newtheorem{question}{Question}[section]
\newtheorem{coro}{Corollary}[section]

\newcommand{\beq}{\begin{equation}}
\newcommand{\eeq}{\end{equation}}
\newcommand{\bea}{\begin{array}}
\newcommand{\ena}{\end{array}}
\newcommand{\bds}{\begin {itemize}}
\newcommand{\eds}{\end {itemize}}
\newcommand{\bdf}{\begin{definition}}
\newcommand{\blm}{\begin{lemma}}
\newcommand{\edf}{\end{definition}}
\newcommand{\elm}{\end{lemma}}
\newcommand{\bthm}{\begin{theorem}}
\newcommand{\ethm}{\end{theorem}}
\newcommand{\bprp}{\begin{prop}}
\newcommand{\eprp}{\end{prop}}
\newcommand{\bcl}{\begin{claim}}
\newcommand{\ecl}{\end{claim}}
\newcommand{\bcr}{\begin{coro}}
\newcommand{\ecr}{\end{coro}}
\newcommand{\bquest}{\begin{question}}
\newcommand{\equest}{\end{question}}

\newcommand{\rarrow}{{\rightarrow}}

\begin{document}
\title{Game theoretic aspects of distributed spectral coordination
with application to DSL networks}
\author{Amir Laufer$^{1,2}$, Amir Leshem$^1$ and Hagit Messer$^2$
\thanks{$^1$ School of Engineering, Bar-Ilan university, Ramat-Gan, 52900,
Israel. $^2$ Department of EE, Tel Aviv
University. This research was funded by the EU-FP6 U-BROAD project
under contract no. 506790. Contact Author: Amir Leshem, e-mail:
leshema@eng.biu.ac.il Part of this paper have been presented at DySpan
2005.}}
\date{}
\maketitle
\begin{abstract}
In this paper we use game theoretic techniques to study the value of
cooperation in distributed spectrum management problems. We show
that the celebrated iterative water-filling algorithm is subject to
the prisoner's dilemma and therefore can lead to severe degradation
of the achievable rate region in an interference channel
environment. We also provide thorough analysis of a simple two bands
near-far situation where we are able to provide closed form tight
bounds on the rate region of both fixed margin iterative water filling
(FM-IWF) and dynamic frequency division multiplexing (DFDM)
methods. This is the only case where such analytic expressions are
known and all previous studies included only
simulated results of the rate region.
We then propose an alternative algorithm that
alleviates some of the drawbacks of the IWF algorithm in near-far
scenarios relevant to DSL access networks. We also provide
experimental analysis based on measured DSL channels of both
algorithms as well as the centralized optimum spectrum management.

\end{abstract}
\section{Introduction}
Recent years have shown great advances in digital subscriber line
(DSL) spectrum management. The public telephone copper lines network
is limited by crosstalk between lines. As such dynamic management of
the lines based on the actual crosstalk channels is becoming an
important ingredient in enhancing the overall network performance at
the physical layer. In a series of papers \cite{Song} \cite{yu01},
\cite{yu2002}, \cite{ginis2002} (and the references therein)  Cioffi
and his group defined several levels of spectral coordination for
DSL access networks, where level zero coordination corresponds to no
coordination, level one corresponds to distributed spectrum
coordination, level two is centralized spectrum management where all
spectral allocations are performed by a single spectrum management
center (SMC). The third level is actually joint transmission /
reception of all lines. To perform level three all signals are
vectored into a single vectored signal. DSM level three can be
divided into two types of vectoring: Two sided coordination (where
all lines are both jointly encoded and jointly decoded) and single
sided coordination where a central processing unit at the network
side of the lines jointly encodes all the downstream transmission or
jointly decodes the upstream transmissions.
Two sided coordination is typical to
private networks, and is implemented e.g., in gigabit Ethernet and
the future 10 Gb Ethernet over copper. Single sided level three
coordination is more relevant to public DSL networks where different
lines are terminated at different customer houses. However joint
transmission over all lines in a binder is still computationally
complicated to implement due to several factors. First equipment
already deployed uses the single input single output approach, where
each line is operated independently assuming interference from other
lines to be part of the background noise. Second the unbundling of
the copper infrastructure and the deployment of remote terminals
makes joint transmission impossible in certain scenarios. It is
anticipated that fiber to the basement and fiber to the neighborhood
architecture will benefit greatly from level three coordination,
while legacy DSL deployment will not be enhanced by these
techniques.
On the other hand dynamic spectrum management (DSM) levels 1-2 only the power spectral
density is optimized to enhance overall network performance is still
an important tool for increasing the reach and improving the service
of legacy long loops. The major difference between DSM level 1 and
level 2 is the existence of a central spectrum management center
performing the optimization jointly at level 2, while DSM level 1
requires distributed coordination of the lines, where each modem
performs its optimization independently of the other lines. The most
appealing property of level 1 coordination is the fact that it can be
implemented  using firmware upgrades to existing DSL modems (which
already have a built in power spectral density (PSD) shaping
capability), rather than complete replacement of infrastructure.

The basic approach to distributed coordination has been proposed in
\cite{yu2002}. In this approach each modem is using the
iterative waterfilling (IWF) algorithm to optimize its own spectrum.
The modem iteratively optimizes its own transmit PSD against the
actual noise caused by other modems in the binder. All modems repeat
this process until convergence is achieved. There are three
basic versions of the IWF algorithm \cite{cioffiDSM}: Rate Adaptive
(RA) where the modem uses all the power to maximize the rate, Margin
Adaptive (MA) where excessive power is used to increase the margin
and Fixed Margin (FM) where the modem minimizes the transmit power
subject to a fixed margin and fixed rate constraint. This is done by
reducing the power whenever the margin achieved is higher than
required. This approach leads to great improvement over the totally
selfish strategies of RA-IWF and MA-IWF. However as we shall
demonstrate, large improvements can be achieved when the modems use
a-priori agreed upon cooperative strategy.

Distributed coordination is basically a situation of conflict
between the users. Each user would like to improve its rate even at
the expense of other users. To gain some insight into the problem we
apply game theoretic techniques. The distributed spectrum management
process can be viewed as a game which is called the interference
game \cite{yu2002}. In this game each user has a pay-off function
given by its rate, and its strategies are basically choice of PSD. A
fixed point of the IWF process is a Nash equilibrium in the
interference game. However Nash equilibrium points can be highly
suboptimal due to the well known Prisoner's dilemma \cite{owen}.
This suggests that defining a new cooperative game where players can
commit to follow certain strategies will improve not only the
overall network capacity, but also the individual user capacity (The
payoff in the interference game is the achievable rate or capacity).
A simple case of the interference game is the two users game. While
this game is rather simplistic it captures well the interference
environment between two {\em groups} of users: One group served from
central office (CO) using legacy equipment such as ADSL
or ADSL2+, and a second group served from a remote
terminal (RT) over shorter lines and more modern equipment such as
VDSL2 modems. It can also model well the case of
two remote terminals of different service providers sharing
customers in the same binder. These two cases are of great interest
from practical point of view. Both cases influence the possible
regulation of spectrum in an unbundled binder. Furthermore the case
of remote terminals is crucial for maintaining legacy service
integrity while expanding the network with remote terminals.

The rest of the paper is organized as follows:
Section II formalizes the distributed spectrum coordination for
Gaussian interference channel in terms of
game theory. It is followed by Section III, in which  the occurrence
of the prisoner's dilemma for a simplified symmetric two players game
is analyzed. Section IV is devoted to the application of the previous
results to the  near-far problem in DSL channels. It provides analytic
expression for the  region where frequency division multiplexing will
improve the rate  region over the competitive IWF algorithm. In
Section V we propose a  simple dynamic frequency domain multiplexing
(DFDM) scheme that  can outperform the IWF in these cases. The results
are also  demonstrated on measured VDSL channels provided by France
Telecom research  labs (Section VI).

\section{The Gaussian interference game}
\label{sec:GI_game} In this section we define the Gaussian
interference game, and provide some simplifications for dealing with
discrete frequencies. For a general background on non-cooperative
games we refer the reader to \cite{owen} and \cite{basar82}.
The Gaussian interference game was defined in
\cite{yu2002}. In this paper we use the discrete approximation
game. Let $f_0 < \cdots <f_K$ be an increasing sequence of
frequencies. Let $I_k$ be the closed interval be given by
$I_k=[f_{k-1},f_k]$. We now define the approximate Gaussian
interference game denoted by $GI_{\{I_1, \ldots, I_K\}}$.

Let the players $1,\ldots,N$ operate over separate channels. Assume
that the $N$ channels have crosstalk coupling functions $h_{ij}(k)$.
Assume that user $i$'th is allowed to transmit a total power of
$P_i$. Each player can transmit a power vector $\vp_i=\left(
p_i(1),\ldots,p_i(K) \right)  \in [0,P_i]^K$ such that $p_i(k)$ is
the power transmitted in the interval $I_k$. Therefore we have $
\sum_{k=1}^K p_i(k)=P_i$. The equality follows from the fact that in
non-cooperative scenario all users will use the maximal power they
can use. This implies that the set of power distributions for all
users is a closed convex subset of the cube $\prod_{i=1}^N
[0,P_i]^K$ given by: \beq \label{eq_strategies} \mB=\prod_{i=1}^N
\mB_i \eeq where $\mB_i$ is the set of admissible power
distributions for player $i$ is \beq \mB_i=[0,P_i]^K\cap
\left\{\left(p(1),\ldots,p(K)\right): \sum_{k=1}^K p(k)=P_i \right\}
\eeq Each player chooses a PSD $\vp_i=\left<p_i(k): 1\le k \le N
\right > \in \mB_i$. Let the payoff for user $i$ be given by: \beq
\label{eq_capacity}
C^i\left(\vp_1,\ldots,\vp_N\right)= \\
\sum_{k=1}^{K}\log_2\left(1+\frac{|h_i(k)|^2p_i(k)}{\sum
|h_{ij}(k)|^2 p_j(k)+\vn(k)}\right)
\eeq where $C^i$ is the capacity available to player $i$ given power
distributions $\vp_1,\ldots,\vp_N$, channel responses $h_i(f)$,
crosstalk coupling functions $h_{ij}(k)$ and $n_i(k)>0$ is external
noise present at the $i$'th channel receiver at frequency $k$. In
cases where  $n_i(k)=0$ capacities might become infinite using FDM
strategies, however this is non-physical situation due to the
receiver noise that is always present, even if small. Each $C^i$ is
continuous on all variables.

\begin{definition}
The Gaussian Interference game $GI_{\{I_1,\ldots,I_k\}}=\left\{\mC,\mB\right\}$ is the N
players non-cooperative game with payoff vector
$\mC=\left(C^1,\ldots,C^N \right)$ where $C^i$ are defined in
(\ref{eq_capacity}) and $\mB$ is the strategy set defined by (\ref{eq_strategies}).
\end{definition}

The interference game is a special case of non-cooperative N-persons
game. An important notion in game theory is that of a Nash
equilibrium.
\bdf
An $N$-tuple of strategies $\left<\vp_1,\ldots,\vp_N\right>$ for
players $1,\ldots,N$ respectively
is called a Nash equilibrium iff for all $n$ and for all $\vp$ ($\vp$ a
strategy for player $n$)
\[
C^n\left(\vp_1,...,\vp_{n-1},\vp,\vp_{n+1},\ldots,\vp_N \right)<
C^n\left(\vp_1,...,\vp_{N} \right)
\]
i.e., given that all other players $i \neq n$ use strategies $\vp_i$, player
$n$ best response is $\vp_n$.
\edf
The proof of existence of Nash equilibrium in the general interference
game follows from an easy adaptation of the proof of the this result
for convex games. In appendix A we demonstrate how the continuity of
the joint water-filling strategies is essentially what is needed in order to prove
the existence of Nash equilibrium in the interference game. It is an
adaptation of the result of \cite{nikaido55} as presented in
\cite{basar82}.  An alternative proof relying on differentiability has been given by Chung
et.al \cite{chung2002a}.
A much harder problem is the uniqueness of Nash equilibrium points in
the water-filling game. This is very important to the stability of the
waterfilling strategies. A first result in this direction has been
given in \cite{chung2002}. A more general analysis of the convergence
(although it still does not cover the case of arbitrary channels has
been given in  \cite{luo2005}.

While Nash equilibria are inevitable whenever non-cooperative zero sum
game is played they can lead to substantial loss to all players, compared to a
cooperative strategy in the non-zero sum case. In the next section we demonstrate this
phenomena for a simplified channel model.

\section{The Prisoner's Dilemma for the 2$\times$2 Symmetric Game}
In order to present the benefits of cooperative strategies for
spectral management we first focus on a simplified two users two
frequency bands symmetric game. The channel matrices of this channel
are the follows:
\begin{equation}
\left|H(1)\right|^2=\left[
\begin{array}{cc}
1 & h \\
h & 1 \end{array} \right]   ,    \left|H(2)\right|^2=\left[
\begin{array}{cc}
1 & h \\
h & 1 \end{array} \right]
\end{equation}
where
$H(1)$ and $H(2)$ are the normalized channel matrices for each
frequency band, and
\[
h=|h_{12}(1)|^2=|h_{21}(1)|^2=|h_{12}(2)|^2=|h_{21}(2)|^2
\]
Since in the DSL environment the crosstalk from other user is
smaller than the self channel response (i.e. $h_{ij}(k)<h_i(k)$
$\forall i,j,k$ we'll limit the discussion to $0\leq h < 1$.

In this section we analyze the symmetric $2 \times 2$ interference game and
find the Nash equilibrium which is achieved by both users using the
full spectrum. We then provide full characterization of channel-SNR pairs for which
IWF is optimal as well as full conditions for the two
other situations: (in terms of pairs of channel coefficient and SNR)
The first is known as the Prisoner's dilemma (PD) and was discovered by Flood and
Dresher \cite{flood52}. The second is the ``chicken'' dilemma game, a
termed coined by B. Russel in the context of the missile crisis in
Cuba \cite{poundstone92}. We will
show that in both these cases cooperative strategies (FDM) outperform the Nash
equilibrium achieved by the IWF.

In our symmetric game both users have the same power constraint $P$ and the power
allocation matrix is defined as
\begin{equation}P\cdot \left[
\begin{array}{cc}
1-\alpha & \alpha \\
\beta & 1-\beta
\end{array} \right]
\end{equation}
The capacity for user I is as follows:
\begin{equation}
C^1=\frac{1}{2}\log_2\left(1+\frac{(1-\alpha)\cdot P}{N+\beta\cdot
P\cdot h}\right)+\frac{1}{2}\log_2\left(1+\frac{\alpha\cdot
P}{N+(1-\beta)\cdot P\cdot h}\right)
\end{equation}\\
where $N$ is the noise power spectral density.\\
The last equation can be rewritten as -
\begin{equation}
C^1=\frac{1}{2}\log_2\left(1+\frac{(1-\alpha)}{SNR^{-1}+\beta\cdot
h}\right)+\frac{1}{2}\log_2\left(1+\frac{\alpha}{SNR^{-1}+(1-\beta)\cdot
h}\right)
\end{equation}
where $SNR = P/N$.\\
By the definition of the Gaussian interference game, the set of
strategies in this simplified game is \beq \label{eq_2x2_strategy}
 \left\{\alpha, \beta :
0\leq \alpha, \beta \leq 1\right\} \eeq
\begin{claim}
\label{theorem_nash} In the $2\times 2$ symmetric interference game
there is Nash equilibrium point at $\alpha = \beta = \frac{1}{2}$.
\end{claim}
\begin{proof}
An IWF solution for this case will be of the form:
\begin{equation}
\label{iwf_eq1}
 (1-\beta_{i-1})h+\alpha_{i}=\beta_{i-1}
h+(1-\alpha_{i})
\end{equation}
\begin{equation}
\label{iwf_eq2} (1-\alpha_{i-1})h+\beta_{i}=\alpha_{i-1}
h+(1-\beta_{i})
\end{equation}
which implies that
\begin{equation}
\label{iwf_eq3} \alpha_{i}=\frac{(2\beta_{i-1}-1)h+1}{2}
\end{equation}
\begin{equation}
\label{iwf_eq4} \beta_{i}=\frac{(2\alpha_{i-1}-1)h+1}{2}
\end{equation}
The expression in (\ref{iwf_eq1}) is the water filling solution for
$\alpha$ in the $i^{th}$ iteration of the IWF as a function of
$\beta$ computed in the $(i-1)^{th}$ iteration. Similarly
(\ref{iwf_eq2}) is the water filling solution for $\beta$ in the
$i^{th}$ iteration as a function of $\alpha$ computed in the
$(i-1)^{th}$ iteration. These set of equations will converges when
\begin{equation}
\label{converges_eq1} \alpha_i = \alpha_{i-1} \equiv \alpha
\end{equation}
and
\begin{equation}
\label{converges_eq2} \beta_i = \beta_{i-1} \equiv \beta
\end{equation}
substituting (\ref{converges_eq1}) and (\ref{converges_eq2}) in
(\ref{iwf_eq3}) and (\ref{iwf_eq4}) and solving the two equations we
get
\begin{equation}
\alpha = \beta = \frac{1}{2}
\end{equation}
since the IWF converges to a Nash equilibrium we conclude
that $\alpha = \beta = \frac{1}{2}$ is a Nash equilibrium in this
game.
\end{proof}
We interpret the IWF as the competitive act, since each user
maximizes its rate given the other user power allocation, we choose
FDM as the cooperative way. Applying FDM (which implies that
$\alpha=\beta=0$) means causing no interference to the other user
,by using orthogonal bands for transmission. We want to compare
between these two approaches of power allocation, the competitive
one (IWF) and the cooperative one (FDM). Instead of comparing these
approaches on the "continuous" game (continuous with respect to the
set of strategies in the game defined in
(\ref{eq_2x2_strategy})), we can discuss and analyze the "discrete"
game, which is characterized by having only two strategies followed by a
set of four different values of $\alpha$ and $\beta$. This
reduction is allowed since for two strategies and two users there
are four different choices of mutual power allocations:
\begin{itemize}
\item both users select FDM resulting in $\alpha=\beta=0$
\item user I selects FDM while user II selects IWF resulting in $\alpha=0$ ,
$\beta=\frac{1-h}{2}$ ($\beta$ is the solution of \ref{iwf_eq4}
where $\alpha=0$)
\item user I selects IWF while user II selects FDM  resulting in $\alpha=\frac{1-h}{2}$ , $\beta=0$ ($\alpha$ is the solution of \ref{iwf_eq3}
where $\beta=0$)
\item both users select IWF resulting in $\alpha=\beta=\frac{1}{2}$ as we have shown
in the theorem
\end{itemize}
Tables I describes the payoffs of users I at four different levels
of mutual cooperation (The payoffs of user II are the same with the
inversion of the cooperative/competetive roles).
\begin{table}[htbp]
\centering \caption{User I payoffs at different levels of mutual
cooperation}
\begin{tabular}{|c|c|c|}
\hline
&user II is fully cooperative&user II is fully competing\\
&$(\beta=0)$&$\left(\beta=\frac{(2\alpha-1)h+1}{2}\right)$\\
\hline $\begin{array}{c}
$user I is fully cooperative$ \\
(\alpha=0)
\end{array}$ & $\frac{1}{2}\log_2\left(1+\frac{1}{SNR^{-1}}\right)$&$\frac{1}{2}\log_2\left(1+\frac{1}{SNR^{-1}+\frac{(1-h)}{2}h}\right)$\\
\hline $\begin{array}{c}
$user I is fully competing$ \\
\left(\alpha=\frac{(2\beta-1)h+1}{2}\right)\\
\end{array}$ & $\frac{1}{2}\log_2\left(1+\frac{\frac{1+h}{2}}{SNR^{-1}}\right)+\frac{1}{2}\log_2\left(1+\frac{\frac{1-h}{2}}{SNR^{-1}+h}\right)$&$\log_2\left(1+\frac{\frac{1}{2}}{SNR^{-1}+\frac{1}{2}h}\right)$\\
\hline
\end{tabular}
\label{Table1}
\end{table}


For certain values of the payoff (determined by channel and SNR) in the
interference game it might be the case that each user can benefit from
other players cooperation, and benefit even more from mutual cooperation. However it
is always the case that given cooperative strategy for the other
player he always benefits from
noncooperation with the others due to the water filling optimality (i.e. given the
interference and noise PSD the best way to allocate the power is
through water filling which, as before mentioned, don't take into
account the influence on other users thus cannot be considered as a
cooperative method). In this situation the stable equilibrium is the
mutual non-cooperation. If on the other hand mutual cooperation is
better for both users over mutual competition we obtain that the
stable point is suboptimal for both players.  This is a well known situation in game theory
 termed the Prisoner's Dilemma \cite{owen} (here and after
abbreviated PD). For a popular overview of the prisoner's dilemma as
well as other basic notions in game theory as well as history of the
subject we recommend \cite{poundstone92}.

A PD situation is defined by the following payoff
relations - $T>R>P>N$, where:

\begin{itemize}
\item $T$ (Temptation) is one's payoff for defecting while the other
cooperates. In our game choosing an IWF while the other player uses FDM.
\item $R$ (Reward) is the payoff of each player where both
cooperate or mutual choice of FDM.
\item $P$ (Penalty) is the payoff of each player when both
defects or mutual use of IWF.
\item $N$ (Naive) is one's payoff for cooperating while the other
defects, i.e., the result of using FDM when the other player uses IWF.
\end{itemize}
It is easy to show that the Nash equilibrium point in this case is
that both players will defect ($P$). This is caused by the fact that
given the other user act the best response will be to defect (since
$T>R$ and $P>N$). Obviously a better strategy (which makes this game
a dilemma) is mutual cooperation (since $R>P$).\\

In our symmetric interference game $\alpha$ and $\beta$ can be
viewed as the level of mutual cooperation. $\alpha$ determines the
level and cooperation of user I with user II, and $\beta$ the level
of cooperation of user II with user I. For analyzing this game we
can analyze the simplified discrete game.  As before mentioned a PD situation is
characterized by the following payoff relations: $T > R > P > N$. By
examining the relations between the different rates (payoffs) as
depicted in table I we can derive a set of conditions on $h$ and
$SNR$ for which the given symmetric interference channel game
defines a PD situation: \\
(a) $T>R$:
\begin{equation}
\frac{1}{2}\log_2\left(1+\frac{\frac{1+h}{2}}{SNR^{-1}}\right)+\frac{1}{2}\log_2\left(1+\frac{\frac{1-h}{2}}{SNR^{-1}+h}\right)>\frac{1}{2}\log_2\left(1+\frac{1}{SNR^{-1}}\right)
\end{equation}
this equation reduces to $h^2-2\cdot h+1>0$ which holds for every
$ h \neq 1$.\\

(b) $T>P$:
\begin{equation}
\frac{1}{2}\log_2\left(1+\frac{\frac{1+h}{2}}{SNR^{-1}}\right)+\frac{1}{2}\log_2\left(1+\frac{\frac{1-h}{2}}{SNR^{-1}+h}\right)>\log_2\left(1+\frac{\frac{1}{2}}{SNR^{-1}+\frac{1}{2}h}\right)
\end{equation}
simplifying the equation we obtain
\begin{equation}
SNR^{-2}\left(h+\frac{1}{4}h^2\right)+SNR^{-1}\left(\frac{1}{2}h^3+\frac{3}{4}h^2+\frac{1}{4}h\right)+\left(\frac{1}{16}h^4+\frac{1}{8}h^3+\frac{1}{16}h^2\right)>0
\end{equation}
since $SNR$ and $h$ are nonnegative the equation always true.\\

(c) $R>P$
\begin{equation}
\frac{1}{2}\log_2\left(1+\frac{1}{SNR^{-1}}\right)>\log_2\left(1+\frac{\frac{1}{2}}{SNR^{-1}+\frac{1}{2}h}\right)
\end{equation}
simplifying (15) we get
\begin{equation}
h^2+2hSNR^{-1}-SNR^{-1}>0
\end{equation}
since $h$ is nonnegative the equation holds for $h>h_{\lim1}$, where
\begin{equation}
\label{eq_hlim1}
h_{\lim1}=SNR^{-1}\left(\sqrt{1+\frac{1}{SNR^{-1}}}-1\right)
\end{equation}\\
(d) $R>N$
\begin{equation}
\frac{1}{2}\log_2\left(1+\frac{1}{SNR^{-1}}\right)>\frac{1}{2}\log_2\left(1+\frac{1}{SNR^{-1}+\frac{(1-h)}{2}h}\right)
\end{equation}
which reduces to $\frac{1-h}{2}\cdot h>0$, this equation holds for every $0\leq h < 1$.\\
(e) $P>N$
\begin{equation}
\log_2\left(1+\frac{\frac{1}{2}}{SNR^{-1}+\frac{1}{2}h}\right)>\frac{1}{2}\log_2\left(1+\frac{1}{SNR^{-1}+\frac{(1-h)}{2}h}\right)
\end{equation}
or equivalently
\begin{equation}
\label{eq_p>n}
 h^3+h^2(0.5+2SNR^{-1})-0.5h-SNR^{-1}<0
\end{equation}
since $h$ is nonnegative the equation holds for $h<h_{\lim2}$, where
$h_{\lim2}$ is the solution for (\ref{eq_p>n}) given by the cubic
formula.
Another condition arises from the sum-rate perspective is the
following - $2R > T+N$. This condition implies that a mixed strategy
(i.e. one user is cooperating while the other competing)
will not achieve higher sum rate than mutual cooperation - \\
(f) $2R > T+N$:
\begin{equation}
\log_2\left(1+\frac{1}{SNR^{-1}}\right)>\frac{1}{2}\log_2\left(1+\frac{\frac{1+h}{2}}{SNR^{-1}}\right)+\frac{1}{2}\log_2\left(1+\frac{\frac{1-h}{2}}{SNR^{-1}+h}\right)+\frac{1}{2}\log_2\left(1+\frac{1}{SNR^{-1}+\frac{(1-h)}{2}h}\right)
\end{equation}
which reduced to
\begin{equation}
SNR^{-2}\left(6(1-h^2)+8h\right)+SNR^{-1}(9h+h^2)+4h^2(1-h)>0
\end{equation}
since $h$ and $SNR$ are nonnegative the equation is true in the
relevant region of $0\leq h<1$ for every $SNR$.
Combining all the relation above we conclude that only three
situation are possible:
\begin{itemize}
\item (A)  $T>P>R>N$ , for $h<h_{\lim1}$
\item (B)  $T>R>P>N$ , for $h_{\lim1}<h<h_{\lim2}$
\item (C)  $T>R>N>P$ , for $h_{\lim2}<h$
\end{itemize}
where $h_{\lim1}$ and $h_{\lim2}$ are given above.\\
The sum rate is either $2\cdot R$ (when both applying FDM), $2\cdot
P$ (when both using IWF) or $T+N$ (when one uses IWF while the other
applying FDM). Examining the achieved sum rate for the two
strategies (IWF and FDM) yields
the following:
The payoff relations in (A) corresponds to a game called
"Deadlock". In this game there is no dilemma, since as in the PD
situation, no matter what the other player does, it is better to
defect ($T>R$ and $P>N$), so the Nash equilibrium point is $P$.
However in contrast to PD, in this game $P>R$ thus there is no reason to
cooperate. The maximum sum rate is also $P$ because $2\cdot R>T+N$
and $P>R$. Since applying the IWF strategy equals to $P$ (by our
definition of competition), this is the region where the IWF
algorithm achieves the maximum sum rate as well as optimal rate for
each user.

The payoff relations in (B) corresponds to the above discussed PD
situation. While the Nash equilibrium point is $P$, the maximum sum
rate is achieved by $R$. In this region the FDM strategy will
achieve the maximum sum rate.

The last payoff relations (C) corresponds to a game called
"Chicken". This game has two distinguished Nash
equilibrium points, $T$ and $N$. This is caused by the fact that for
each of the other player's strategies the opposite response is preferred
(if the other cooperates it is better to defect since $T>R$, while
if the other defects it is better to cooperate since $N>P$). The
maximum rate sum point is still at $R$ (since $R>P$ and $2\cdot
R>T+N$) thus, again FDM will achieve the maximum rate sum while IWF
will not.

An algorithm for distributed power allocation can be derived from
this insight for the symmetric interference game. Given a symmetric
interference game (i.e. a symmetric channel matrix and $SNR$), if
$h<h_{\lim1}$ (where $h_{\lim1}$ is given in (\ref{eq_hlim1})) use
the IWF method to allocate the power, else, both players should use
the FDM method. Since the channel crosstalk coefficient $h$ is
assumed to be known to both users this algorithm can be implemented
distributively (with pre agrement on the band used by each user for
the FDM). We will return to this strategy in the context of real DSL
channels in section \ref{DFDM}

It is important to distinguish between the  continuous symmetric
interference game and the discrete one. Even though the discrete
game can have Nash equilibrium other than $\alpha = \beta =
\frac{1}{2}$ (as we saw in the chicken game) these equilibrium
points are not stable in the continuous game. Hence we are left with only one
stable equilibrium as proven in (\ref{theorem_nash}). Nevertheless,
our conclusions regarding the benefit of cooperation in the interference game derived from the
discrete game remains valid in the continuous one since once
continuous strategies are chosen they inevitably lead to
$\ga=\gb=0.5$. However when players choose to cooperate the stability issue
is not important since IWF is not used.
\begin{figure}
\begin{center}
   \mbox{\psfig{figure=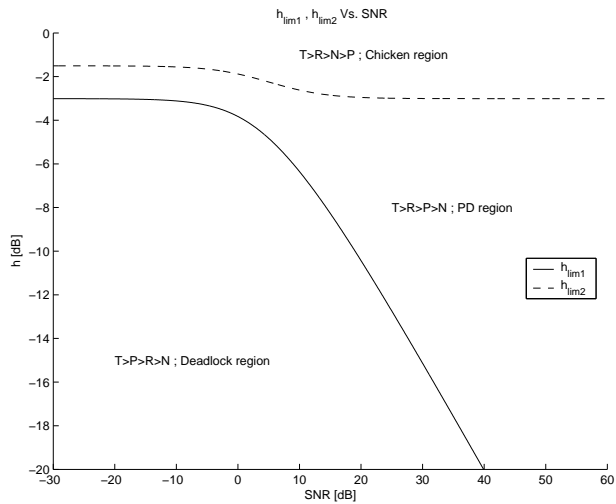,width=0.45\textwidth}}
\end{center}
   \caption{Graph of $h_{\lim1}$ , $h_{\lim2}$ Vs. SNR, The solid line
corresponds to $h_{\lim1}$ and the dashed line corresponds to
$h_{\lim2}$} \label{fig 1}
\end{figure}
Further discussion and examples of the prisoner's dilemma in this case
can be found in \cite{laufer2005}.
\section{The Near-Far Problem}
One of the most important spectral coordination problems in the DSL
environment is the near-far problem. This problem has similarity to
the power control problem in CDMA network. However the DSL channel
is frequency selective (see Figure \ref{figure_H}) and multi-carrier modulation is typically
used. Therefore the interference from remote terminal to CO based
services is very severe and has properties similar to near end
crosstalk (NEXT). This scenario is typical to unbundled loop plants
where the incumbent operator is mandated by law to lease CO based
lines to competitive  operators. Figure \ref{near_far_topology}
describes a typical near far interference environment.

The problem has also appeared in the upstream direction of VDSL
(which is at frequencies above 3MHz). The solution of the VDSL
standard is highly suboptimal since the optimization has been done
for fixed services under specific noise scenarios. It has been shown
that upstream spectral coordination can lead to significant
enhancement of upstream rates in real life environments. While DSL
channels have relatively complicated frequency response and full
analysis is possible only based on computer simulations and measured
channels, we provide here an analysis of a simplified near far
scenario that captures the essence of distributed cooperation in
near far scenarios. In section \ref{simulations} we will provide
simulated experiments on measured channels.

The analysis in this section is divided into two parts. First a
simple symmetric bandwidth near-far game with no option to partition
the bands is analyzed and it is proved that an FDM solution is
optimal. Then the results are extended to a more general situation
with asymmetric bandwidth. In this case we show that a solution
minimizing the interference by utilizing only part of the band is
preferable to a global FM-IWF. This is done by providing analytic
bounds on the rate region for both strategies. Unlike all previous
analysis of these strategies we are able to provide analytic bounds
on the rate region.
\subsection{Symmetric two bands Near-Far problem}
Consider the case of two users using two bands with channel matrices
given by
\begin{equation}
\label{ISI_channel}
\left|H(1)\right|^2=\left[
\begin{array}{cc}
\ga & \gb \\
\grg & 1 \end{array} \right]   ,
\left|H(2)\right|^2=\left[
\begin{array}{cc}
0 & \gd \\
\gre & 1 \end{array} \right]
\end{equation}
where
$H(1)$ and $H(2)$ are the normalized channel matrices for each
frequency band. Note that the second band can be used only by the
second user which will be termed the strong user. Furthermore we
assume that the first band can be used partly by the second user if he
chooses a non-naive FDM or non-naive TDM strategies.
The first user will be called
the weak user.
To simplify the discussion we make the following assumptions:
\bds
\item Both users have transmit power limitation $P$. This is not
essential but simplifies notation.
\item $\ga<<1$ This is the reason that we refer to the first user as
the weak user.
\item $N_i$ is the additive Gaussian noise is constant for both
receivers and at both bands. This assumption is reasonable since the
design of all multi-carrier modems requires low modem noise
floor in order to support the high constellations.
\item $N_i<<\gb P$ This means that the weak user is limited by the
crosstalk from the strong user.
\item $\grg P<<N_i<<0.5 P$. Typically the weak lines emerging from the CO
generate crosstalk that is negligible into the RT line. This means
that basically the strong user sees the same signal to noise ratio
across the two bands. This is actually better for the weak user than
the real situation where the strong user observes {\em better} SNR
on the first band. The second inequality suggest that we work in the
bandwidth limited high SNR regime, which is the interesting case for
DSL networks.
\item User II can perform a voluntary power backoff $\gt$.
\eds
Under our assumptions user II completely dominates the
achievable rate of user I, and user I has no way to force anything on
user II. This type of game is called ``The Bully'' game, where the
strong user can decide to behave in any manner. We would like to
analyze the benefits of a ``polite bully'' that takes whatever it
needs, but behaves as polite as possible to other users, by allowing them to
use resources he does not need.

To that end we analyze the capacity region of the two users under
water-filling strategies and under interference minimization strategy
of the second user, where
the strong user utilizes only partially the joint resource which is
the first band. Note that all the strategies are purely distributed
since only the agreement to behave politely by the bully player is required.
We make several observations regarding the possible strategies:
\bcl
The weak user will always use all its power in the first band.
\ecl
This claim follows from the fact that user I has no capacity in the
second band.

Let the power allocation of user II be $\left(P_1,P_2 \right)$ such
that $P_1+P_2=P$.
\bcl
The rate achievable by user I is given by
\[
C^1=\log_2\left(1+\frac{\ga P}{\gb P_1+N_1}\right)
\]
\ecl This claim is implied by our assumptions of Gaussian signalling
by both users and independent detection of each user. Typically for
the DSL interference channel, the interference to AWGN ratio is
insufficient for successive interference cancelation so each user
should treat the other users interference as Gaussian noise. It is
now easy to compute the optimal rate adaptive strategy for user II.
\begin{claim}
\label{bully_power_cl}
The power allocation for user II under
politeness factor $\tau$ is given by
\begin{equation}
\label{power_IWF}
\begin{array}{lcl}
P_1 = \frac{P}{2}(\tau-\gamma) & \qquad & P_2 =
\frac{P}{2}(\tau+\gamma)
\end{array}
\end{equation}
\end{claim}
The proof for claim \ref{bully_power_cl} follows the same lines as
the proof of claim \ref{theorem_nash}. The WF solution suggests a
constant level of the transmitted power + noise (which includes the
interference) for each band. In our case this implies that
\begin{equation}
P_1+N_2+\gamma P = P_2+N_2
\end{equation}
since $P_1+P_2=\tau P$ we can rewrite the equation as
\begin{equation}
2P_1+\gamma P=\tau P.
\end{equation}
Solving for $P_1$ we obtain
$P_1 = \frac{P}{2}(\tau-\gamma)$ and $P_2 = \frac{P}{2}(\tau+\gamma)$
We now obtain the rate for user II.
\begin{claim}
The rate of user II under FM-IWF with power backoff $\tau$
is given by:
\begin{equation}
C^2=\log_2\left(1+\frac{P_1}{N_2+\gamma P}\right)+\log_2\left(1+\frac{P_2}{N_2}\right)
\end{equation}
where $P_1$ and $P_2$ are given by (\ref{power_IWF}).
\end{claim}

An alternative approach for user II can be to minimize the
interference to the first band by increasing the
power in the second band should it find it useful. This leads to different
expression for the capacity region.

The expression for the capacities using the cooperative act of
user II have the same form as before
\begin{equation}
\begin{array}{c}
C^1=\log_2\left(1+\frac{\alpha P}{N_1+\beta \tilde{P_1}}\right)\\
C^2=\log_2\left(1+\frac{\tilde{P_1}}{N_2+\gamma
P}\right)+\log_2\left(1+\frac{\tilde{P_2}}{N_2}\right)
\end{array}
\end{equation}
Where $(\tilde{P_1},\tilde{P_2})$ is the new power allocation of
user II such that $\tilde{P_1}+\tilde{P_2}=\tilde{\tau} P$

In order to find the $(\tilde{P_1},\tilde{P_2})$ we need to choose
the minimal $\tilde{P_1}$ such that the following equation holds
\begin{equation}
\label{fdm_eq} \log_2\left(1+\frac{P_1}{N_2+\gamma
P}\right)+\log_2\left(1+\frac{P_2}{N_2}\right)=\log_2\left(1+\frac{\tilde{P_1}}{N_2+\gamma
P}\right)+\log_2\left(1+\frac{\tilde{P_2}}{N_2}\right)
\end{equation}
where $P_1,P_2$ are defined by (\ref{power_IWF}).
It is clear that in order to minimize $\tilde{P_1}$ we need to
set $\tilde{\tau}$ to 1. By doing so we enable user II to allocate the maximum
amount of power on the second band and therefor minimize the power
on the first band. Substituting $\tilde{P_2}$ with $P-\tilde{P_1}$
and solving (\ref{fdm_eq}) for $\tilde{P_1}$ we get
\begin{equation}
\tilde{P_1} = \frac{1}{2}P(1-\gamma)\pm
\frac{1}{2}\left[P(1-\tau)(P+4N_2+2\gamma P+P\tau
)\right]^{\frac{1}{2}}
\end{equation}

Using the minimal solution for $\tilde{P_1}$ and applying some
algebra on the expression above we obtain
\begin{equation}
\tilde{P_1} =
\frac{1}{2}P(1-\gamma)-\frac{1}{2}\left[P^2(1-\tau)^2\left(\frac{1+4N_2/P+2\gamma+\tau}{1-\tau}\right
)\right]^{\frac{1}{2}}
\end{equation}
which can be rewritten as
\begin{equation}
\tilde{P_1} =
\frac{1}{2}P(1-\gamma)-\frac{1}{2}P(1-\tau)\left(\frac{1+4N_2/P+2\gamma+\tau}{1-\tau}\right
)^{\frac{1}{2}}
\end{equation}
Since the term $\left(\frac{1+4N_2/P+2\gamma+\tau}{1-\tau}\right)$
is grater or equal to 1 the square root of this term is also grater
or equal to 1. We can write
\[
1+\Delta=\sqrt{\frac{1+4N_2/P+2\gamma+\tau}{1-\tau}}
\]
where $\Delta$ is some positive constant. Therefore we can
write
\begin{equation}
\label{p1_tilda_eq}
\tilde{P_1} =
\frac{1}{2}P(1-\gamma)-\frac{1}{2}P(1-\tau)(1+\Delta)
\end{equation}
arranging (\ref{p1_tilda_eq}) we get
\begin{equation}
\tilde{P_1} = \frac{1}{2}P(\tau-\gamma)-\frac{1}{2}P(1-\tau)\Delta
\end{equation}
which, by claim \ref{bully_power_cl} becomes
\begin{equation}
\label{final_p1_tilde_eq}
\tilde{P_1} =
P_1-\frac{1}{2}P(1-\tau)\Delta
\end{equation}

If the value of $\tilde{P_1}$ as given in
(\ref{final_p1_tilde_eq}) is negative we should fix it to zero. This
is the best situation for user I as he sees
no interference at all.

Since $C^2$ is equal for both methods (guaranteed by (\ref{fdm_eq}))
and $\tilde{P_1}\leq P_1$ (i.e. the interference that user I sees
using the cooperative method is less than or equal to the one
obtained by FM IWF) we conclude that the rate region achieved using
the cooperative act contains the rate region related to FM IWF.

\subsection{Near-Far problem in the bandwidth limited case}
Our next step will be to extend the analysis above to the case where
the two bands have non-identical bandwidth, and we work in the
bandwidth limited regime, i.e., the spectral efficiency of the
transmission is higher than 1 (we transmit more than 1 bit per
channel use). In this case the signal to noise ratio at each
receiver is positive. This will capture a more realistic ISI limited
channel similar to the DSL channel. We shall restrict the analysis
to flat attenuation in each band.

Assume that the first band has bandwidth $W_1$ and the second band
has bandwidth $W_2$. Similarly to the previous case assume that the
channel matrices at each band are given by (\ref{ISI_channel}).

To simplify the expressions we shall also assume that $\grg P << N_2
W_i$, where $N_2$ is the PSD of the AWGN of the second user
receiver. This is realistic in typical near far problems in DSL
where the FEXT from the CO lines into the RT lines is negligible
compared to the AWGN due to the strong loop attenuation of lines
originating at the CO. Under our assumptions we prove the following:
\bthm \label{RR_IWF} The rate region of the FM-IWF satisfies \beq
W_1 \log_2 \left( 1+\frac{\ga P}{\gb P
\left(2^{\frac{R_2}{W_1+W_2}-1} \right)/{\bar{ \SNR}}_2 +W_1
N_1}\right) \le R_1 \le W_1 \log_2 \left( 1+\frac{\ga P}{\gb P
\left(2^{\frac{R_2}{W_1+W_2}+1} \right)/{\bar{ \SNR}}_2 +W_1
N_1}\right) \eeq where $\gr=\frac{W_1}{W1+W_2}$ and \beq
\label{SNRgeometric} {\bar {\SNR}}_2=\left(\frac{P}{N_2W_1}
\right)^{\gr}\left( \frac{P}{N_2 W_2} \right)^{1-\gr} \eeq is a
generalized geometric mean of the SNR at the two bands. \ethm The
capacity of the two users is now given by\footnote{We will analyze
capacity only so the Shannon gap is $\Gamma=1$ (other gaps can be
treated  similarly with just an extra term $\Gamma$).}:
\begin{equation}
\label{BW_capacuty}
\begin{array}{c}
C^1=W_1\log_2\left(1+\frac{\ga P}{N_1 W_1+\beta {P_1}}\right)\\
C^2=W_1\log_2\left(1+\frac{{P_1}}{N_2 W_1+\gamma
P}\right)+W_2\log_2\left(1+\frac{{P_2}}{N_2 W_2}\right)
\ena
\end{equation}
where again $P_1+P_2=\gt P$.
To determine $\gt$ assume that the target rate of the bully player is
$R_2$ and ignore $\grg P$ by our assumption.
Therefore IWF results in flat transmit PSD for user 2:
\[
\bea{l}
P_1=\gr \gt P \\
P_2=(1-\gr) \gt P
\ena
\]
We require that \beq \label{def_R2} R_2=W_1 \log_2\left(1+\frac{{\gr
\gt P }}{N_2 W_2}\right)+W_2\log_2\left(1+\frac{(1-\gr)\gt P} {N_2
W_2}\right) \eeq Therefore  we obtain \beq R_2 \ge W_1
\log_2\left(\frac{{\gr \gt P }}{N_2
W_1}\right)+W_2\log_2\left(\frac{(1-\gr)\gt P} {N_2 W_2}\right) \eeq
Actually using the high SNR approximation we can replace the
inequality by approximate equality. Hence \beq 2^{R_2} \ge
\left(\frac{\gr \gt P}{N_2 W_1}\right)^{W_1}\left(\frac{(1-\gr) \gt
P}{N_2 W_2}\right)^{W_2} \eeq Hence \beq 2^{\frac{R_2}{W_1+W_2}} \ge
\left(\frac{\gr \gt P}{N_2 W_1}\right)^{\gr}\left(\frac{(1-\gr) \gt
P}{N_2 W_2}\right)^{1-\gr} \eeq Further simplification yields \beq
2^{\frac{R_2}{W_1+W_2}} \ge \frac{\gt P}{N_2 W_1^{\gr} W_2^{1-\gr}}
\gr^{\gr} (1-\gr)^{(1-\gr)} \eeq Therefore \beq
\label{exact_upper_tao} \gt \le
\frac{2^{\frac{R_2}{W_1+W_2}}}{{\bar{\SNR}}_2 \gr^{\gr}
(1-\gr)^{(1-\gr)}} \eeq Also note that since $0 < \gr < 1$
\[
\frac{1}{2} \le \gr^{\gr} (1-\gr)^{(1-\gr)} \le 1
\]
hence \beq \label{tao_bound} \gt \le
\frac{2^{\frac{R_2}{W_1+W_2}+1}} {{\bar{\SNR}}_2} \eeq Substituting
(\ref{tao_bound}) into (\ref{BW_capacuty}) we obtain that the rate
for user I is bounded by \beq R_1 \le W_1 \log_2 \left( 1+\frac{\ga
P}{\gb P \left(2^{\frac{R_2}{W_1+W_2}+1} \right)/{\bar{ \SNR}}_2
+W_1 N_1}\right) \eeq

This is indeed very satisfying. As we know the bully's power backoff
is determined by the required spectral efficiency $R_2/(W_1+W_2)$
and the geometric mean SNR of the bully player. Also note that
no matter how good the SNR of user II on the second band, the FM-IWF
always incurs a loss to user I's capacity, since there is always
additional disturbance in the first band. The total rate can be
rewritten as \beq \label{FM_IWF_rate} R_1 \le W_1 \log_2 \left(
1+\frac{\ga}{\gb \left(2^{R_2/(W_1+W_2)+1}
\right){\bar{\SNR}}_2^{-1}+W_1 \SNR_1^{-1}} \right) \eeq and it is
always lower than the rate of interference free situation. On the
other hand if the rate of user II satisfies
\[
R_2\le W_2 \log_2\left(1+\frac{P}{W_2 N_2} \right)
\]
an FDM strategy will achieve for user I a rate \beq R_1=W_1 \log_2
\left( 1+\frac{\ga}{ W_1 SNR_1^{-1}} \right) \eeq which is always
higher than the right hand side of (\ref{FM_IWF_rate}).

When the signal to noise ratio of user II is positive (BW limited
case) we can also obtain a lower bound on the achievable rate of
user I. Similarly to the previous case we obtain a lower bound on
the rate of user I given a rate $R_2$ for user II. The proof is
similar. Start with (\ref{def_R2}) and note that when \beq
\label{condition} \frac{{\gt P }}{N_2(W_1+ W_2)} \ge 1 \eeq
 we have
\beq \label{upper_boundR2} R_2 \le  W_1 \log_2\left(\frac{{2\gr \gt
P }}{N_2 W_1}\right)+W_2\log_2\left(2\frac{(1-\gr)\gt P} {N_2
W_2}\right) \eeq since $1+x<2x$ for $x>1$ and since
(\ref{condition}) drags $\frac{\gr \gt P}{N_2 W_1} \ge 1$ and
$\frac{(1-\gr) \gt P}{N_2 W_2} \ge 1$. Similar derivation now yields
\beq \label{tao_bound1} \gt \ge \frac{2^{\frac{R_2}{W_1+W_2}-1}}
{{\bar{\SNR}}_2} \eeq Which leads to a lower bound on $R_1$ \beq R_1
\ge W_1 \log_2 \left( 1+\frac{\ga P}{\gb P
\left(2^{\frac{R_2}{W_1+W_2}-1} \right)/{\bar{ \SNR}}_2 +W_1
N_1}\right) \eeq This provides good lower and upper bounds on the
rate region as a function of the channel parameters. As noted for
high SNR scenarios the upper bound on $\gt$ (\ref{exact_upper_tao}) is
tight, which provides accurate estimate
of the rate region. This ends the proof of theorem \ref{RR_IWF}.

We now provide similar bounds on the rate region of a dynamic FDM,
where the bully minimizes the fraction of the first band that he
uses.
\bthm
The rate region of a dynamic FDM strategy where given a rate $R_2$ the strong
player minimizes the fraction of the first band he uses is bounded by
\beq \label{R1_rate_bound}
\bea{lcl}
R_1  & \le &  (1-\gl_{\min}) W_1 \log_2 \left(
1+\frac{\ga P\left(1+\gl_{\min} W_1 \frac{1}{\gl_{\min}
W_1+W_2}\right)}{N_1}\right) \\
     & + & \gl_{\min} W_1\log_2 \left( 1+\frac{\ga
P\left( 1-(1-\gl_{\min}) W_1 \frac{1}{\gl_{\min} W_1+W_2}\right)}{N_1+\gb
\frac{P}{\gl W_1+W_2}}\right) \\
R_1  & \ge & (1-\gl_{\max}) W_1 \log_2 \left(
1+\frac{\ga P\left(1+\gl_{\max} W_1 \frac{1}{\gl_{\max}
W_1+W_2}\right)}{N_1}\right) \\
     & + & \gl_{\max} W_1\log_2 \left( 1+\frac{\ga
P\left( 1-(1-\gl_{\max}) W_1 \frac{1}{\gl_{\max} W_1+W_2}\right)}{N_1+\gb
\frac{P}{\gl W_1+W_2}}\right)
\ena
\eeq
where
\begin{equation}
\label{L_bound}
\bea{lcl}
\gl_{\min} =
\frac{\frac{R_2}{\log_2(1+\frac{SNR_2}{W_2})}-W_2}{W_1}
 & \qquad & \gl_{\max} =
\frac{\frac{R_2}{\log_2(1+\frac{SNR_2}{(W_1+W_2)})}-W_2}{W_1}
\ena
\end{equation}
\ethm The proof of this theorem is given in appendix B.
\section{The dynamic FDM coordination algorithm}
\label{DFDM} 
\begin{figure}[htbp]
\centerline
{
\subfigure[]{\includegraphics[width=6cm]{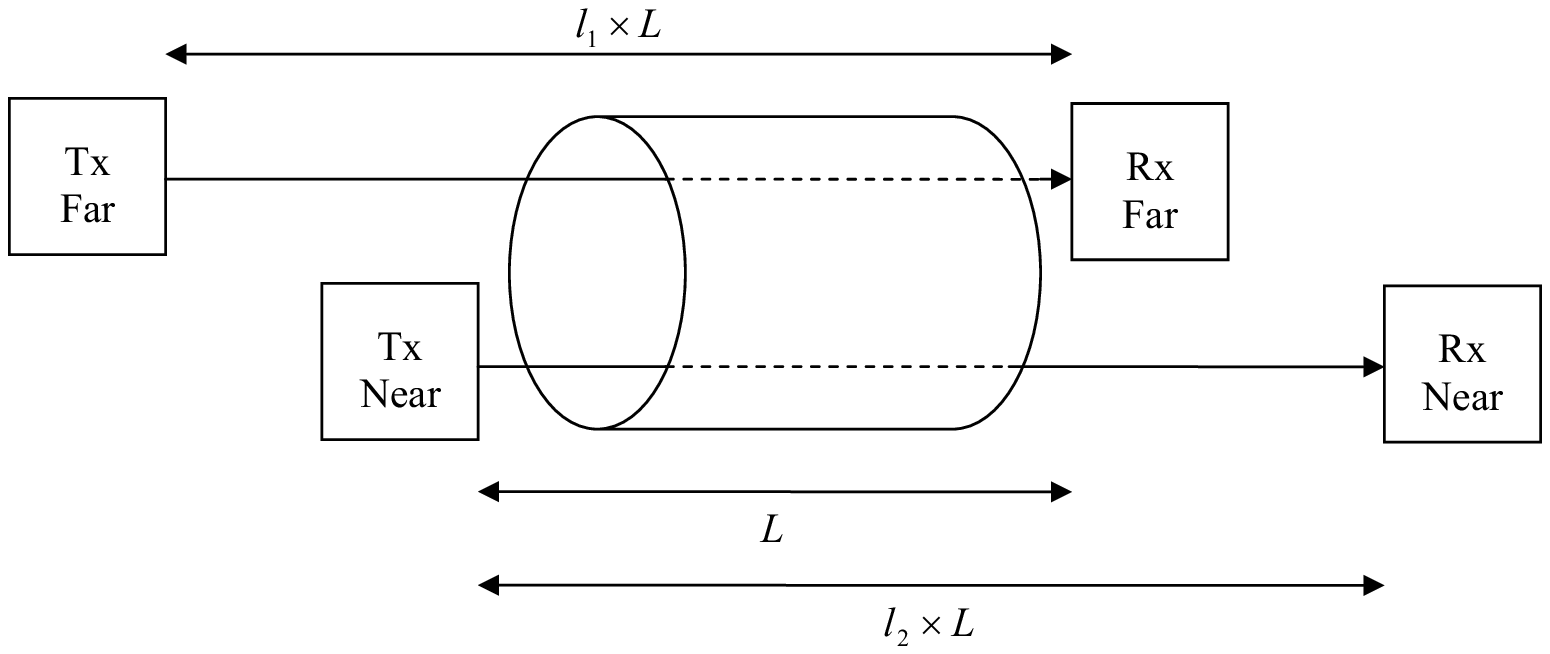}\label{near_far_topology}}\hfil
\subfigure[]{\includegraphics[width=6cm]{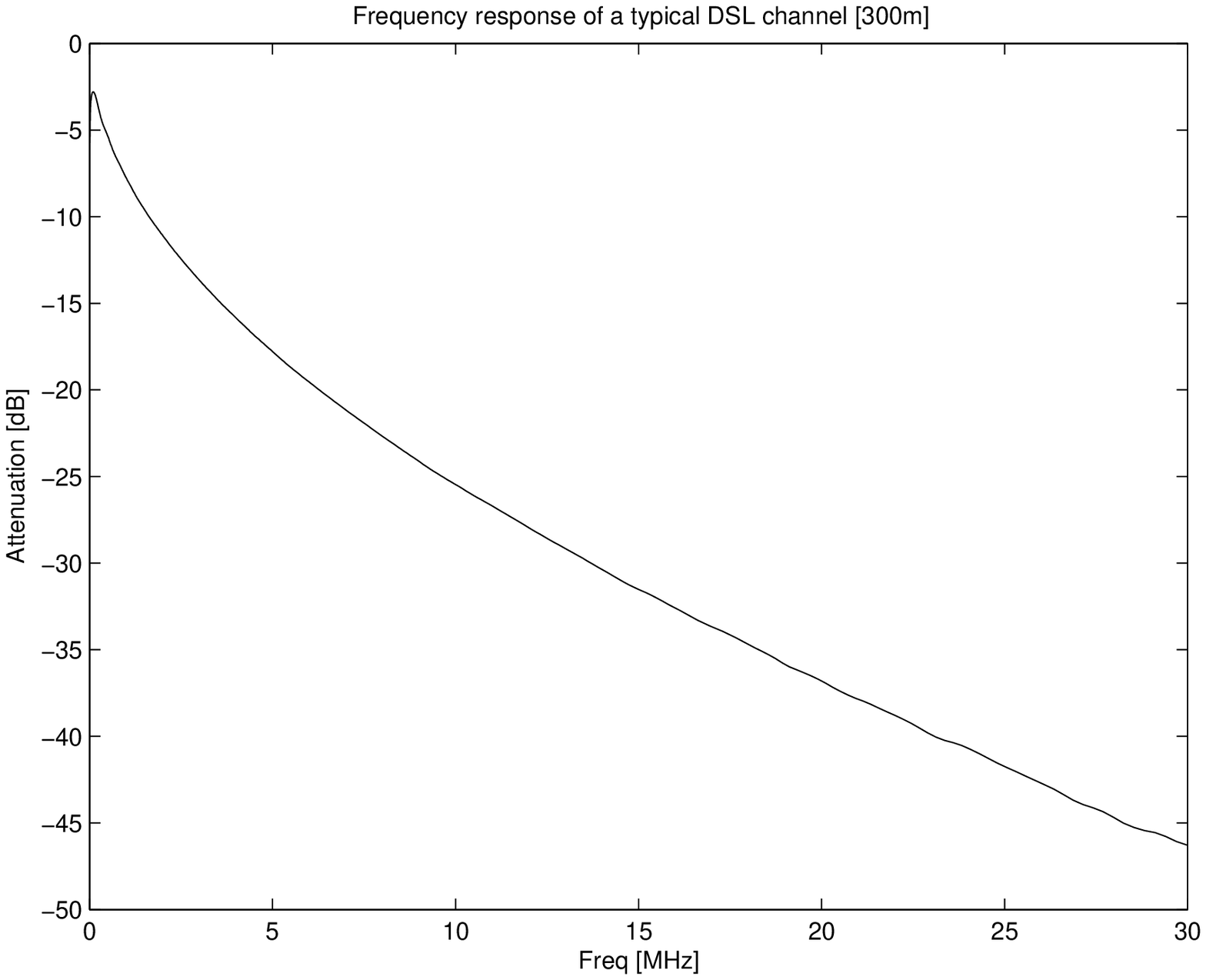}\label{figure_H}}}
\caption{(\ref{near_far_topology}) Loop topology of the Near-Far
problem in DSL. (\ref{figure_H})  Typical VDSL2 channel}
\end{figure}

DSL channels have typically higher attenuation at
higher frequencies. (see figure \ref{figure_H}). A typical DSL topology
including CO and RT deployment is depicted in figure
\ref{near_far_topology}. As we can see the users of the RT are the
Bully type users which do not typically suffer interference from CO
based lines, but do cause substantial interference to the CO based
lines.

Inspired by our analysis of cooperative strategies presented in the
previous sections we propose a cooperative
solution for the near-far problem. The dynamic FDM  (DFDM) algorithm,
first presented in \cite{leshem2003}, allocates the power of the near
user not only as a function of the noise PSD on its own line (as the IWF does) but by
minimizing the use of the lower part of the spectrum. Since the far
user can allocate its power only at the lower part of spectrum, applying the
DFDM on the far user power allocation reduces the level of
interference to the far user by means of orthogonal transmitting
bands. The idea underlying the approach above is that the far user
uses the lower part of the spectrum (as explained above), and
therefore use of this part of the spectrum should be minimized for
the near user. A variation of this method in the centralized level 2
DSM is the band preference method \cite{cioffi2003}.

We define $f_c$ to be the cutoff frequency i.e. the minimal
frequency used by the near user. The power allocation method in the
DFDM algorithm is as follows - given $R_d$ the design rate of the
remote terminal user, the RT user allocates its power such that the
rate achieved is equal to $R_d$ along with maximizing $f_c$. More
precisely the algorithm is implemented in two steps: At the first
step the maximal $f_c$ is found (this step is performed by applying
RA-IWF at varying $f_c$ values). The second step is reducing the
total power by applying FM-IWF on the upper part of the spectrum
determined by the former step. The implementation steps of the DFDM
algorithm are summarized in table II. When the signal to noise ratio
is high we can replace the RA-IWF by computing the capacity based on
the measured noise profile (since for all RT based users the channel
and crosstalk are approximately identical). 
\begin{table}[htbp]
\label{DFDM_tab} \centering \caption{DFDM implementation for the
Near-Far scenario}
\begin{tabular}{|l|}
\hline
1. Let $R_d$ = preassigned target rate for the near user.\\
2. Estimate the received noise PSD.\\
3. find $f_c$, the minimal $f$ such that the near user can achieve
rate $R_d$ using frequencies above $f_c$.\\
4. Allocate the minimal amount of power needed for achieving $R_d$
using only frequencies grater than $f_c$.\\
\hline
\end{tabular}
\end{table}

\section{Simulations}
\label{simulations} 
\begin{figure}[htbp]
\centerline{
\subfigure[]{\includegraphics[width=6cm]{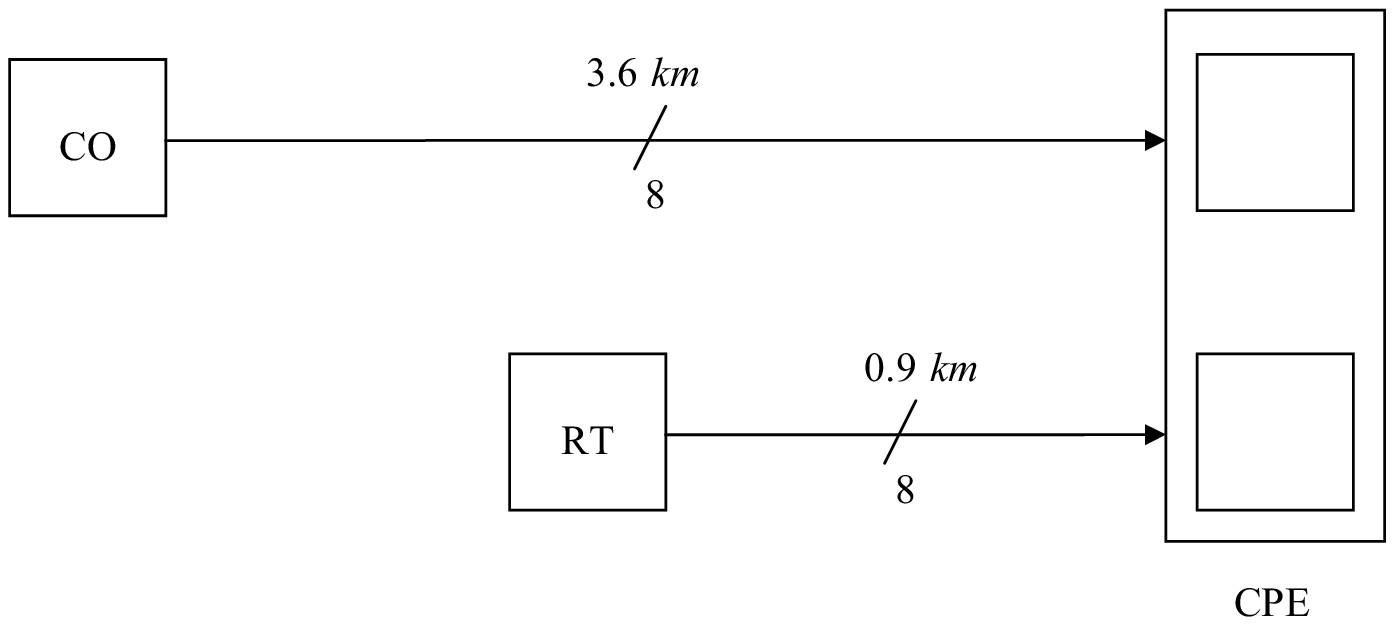}\label{downstream}}
\hfil
\subfigure[]{\includegraphics[width=6cm]{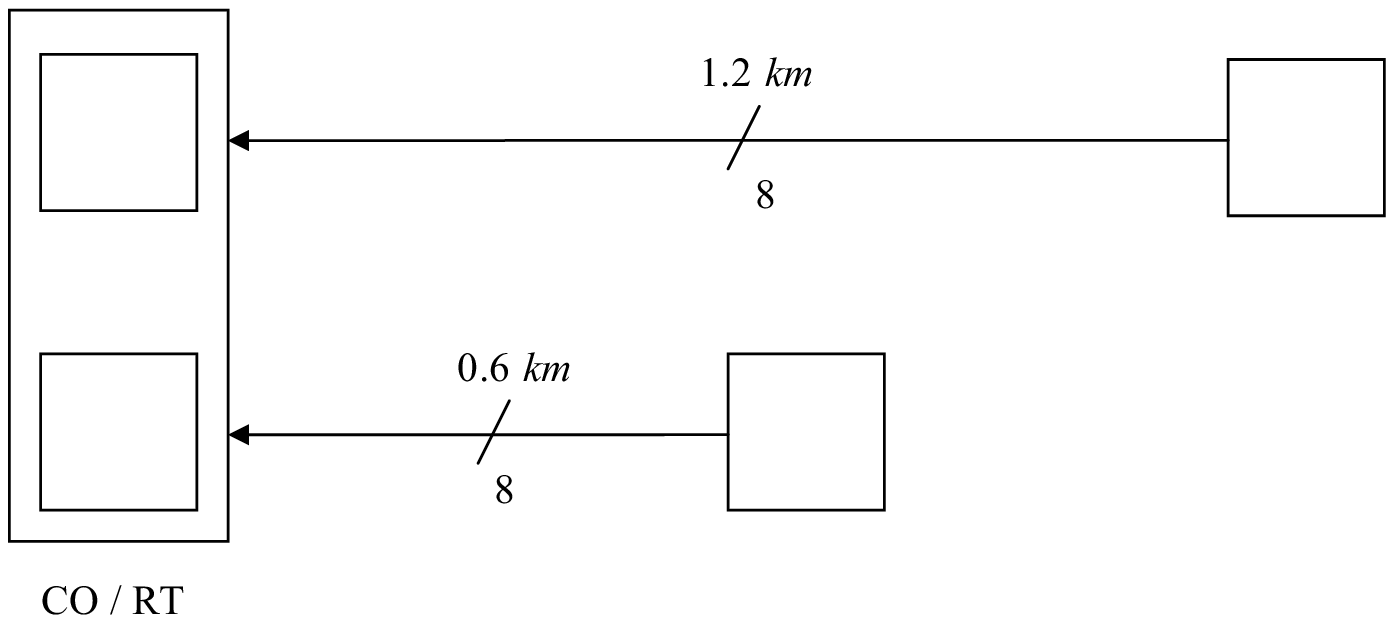}\label{upstream}}}
\caption{(\ref{downstream}) CO/RT downstream setup. (\ref{upstream})Near-Far upstream setup }
\end{figure}

In this section we examine the rate region of
the DFDM algorithm compared to FM-IWF.  We have also simulated the
OSM method \cite{cendrillon}, \cite{cendrillon2005} which is a DSM level 2 in order to have
an upper bound on the performance of DSM level 1 techniques. The
channel transfer matrix is a measured binder provided by France
Telecom research labs \cite{karipidis2005a}. The simulations global
parameters are VDSL 998 band plan up to 12 MHz, a maximum power
constraint of $30mW$ (15 dBm) and a white noise PSD of $-140dBm/Hz$.
In addition the frequency Division Duplex (FDD) 998 bandplan is
used. We have simulate two scenarios: \bds
\item Central office / Remote Terminal Downstream.
\item Upstream with non-identical locations.
\eds
The first scenario represent downstream setup where a central office
(CO) with $8\times 3.6$ km ADSL  lines is sharing a binder with a remote
terminal (RT) with $8\times 0.9$ km VDSL lines. The RT is located
2.7 km from the CO as depicted in Figure \ref{downstream}.
In the second simulation set we have studied upstream coordination. We
have used two clusters of VDSL users sharing the same binder transmitting to the
same RT. The far group contained 8 lines located 1.2 km from the
RT while the near group contained 8 lines located just
600m from the RT, as depicted in Figure \ref{upstream}. Since in the VDSL 998
bandplan the lowest US frequency is 3.75 MHz the near far problem is
much more pronounced than in ADSL.

\begin{figure}[htbp]
\centerline
{
\subfigure[]{\includegraphics[width=6cm]{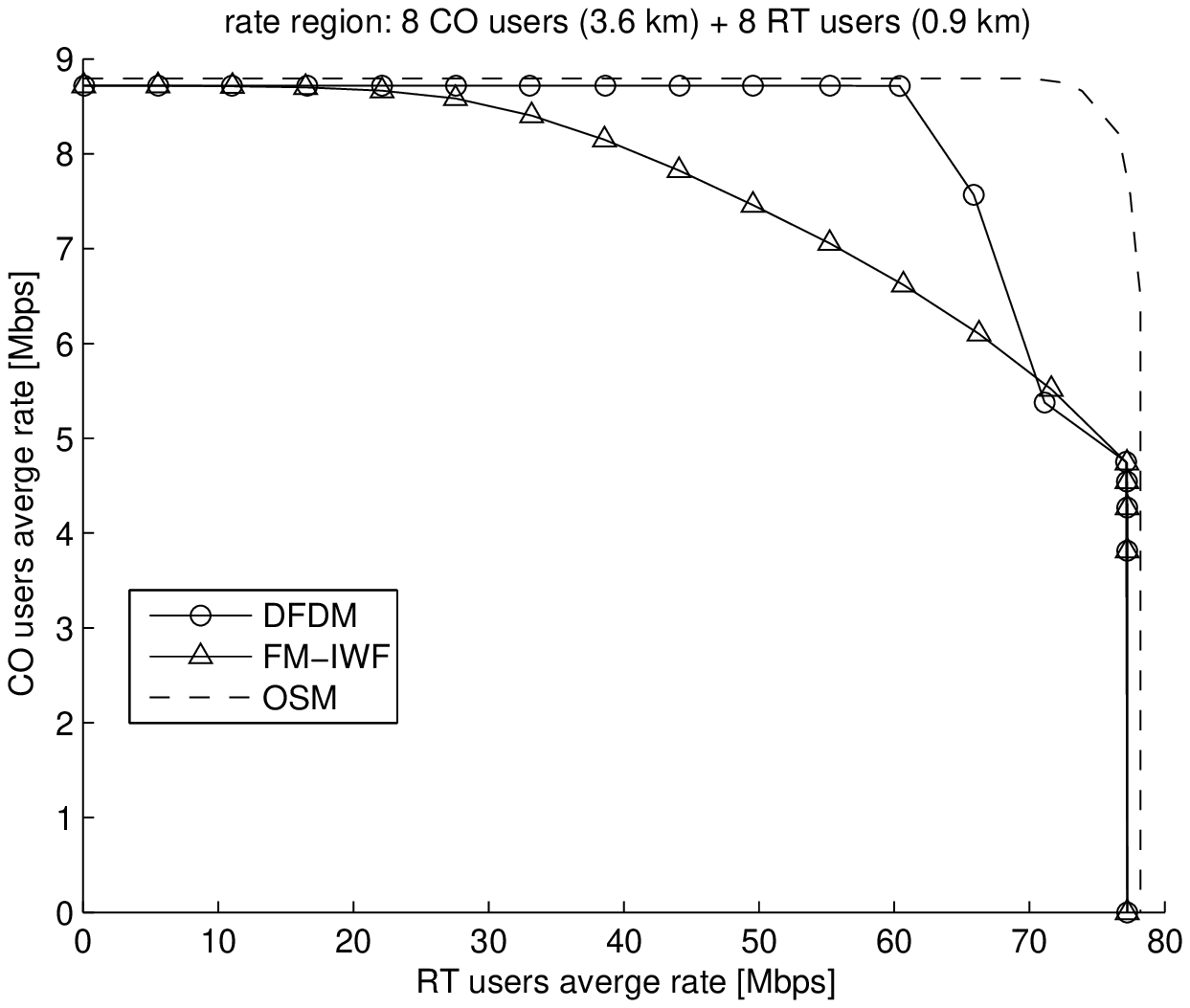}\label{downstream_rr}}
\hfil
\subfigure[]{\includegraphics[width=6cm]{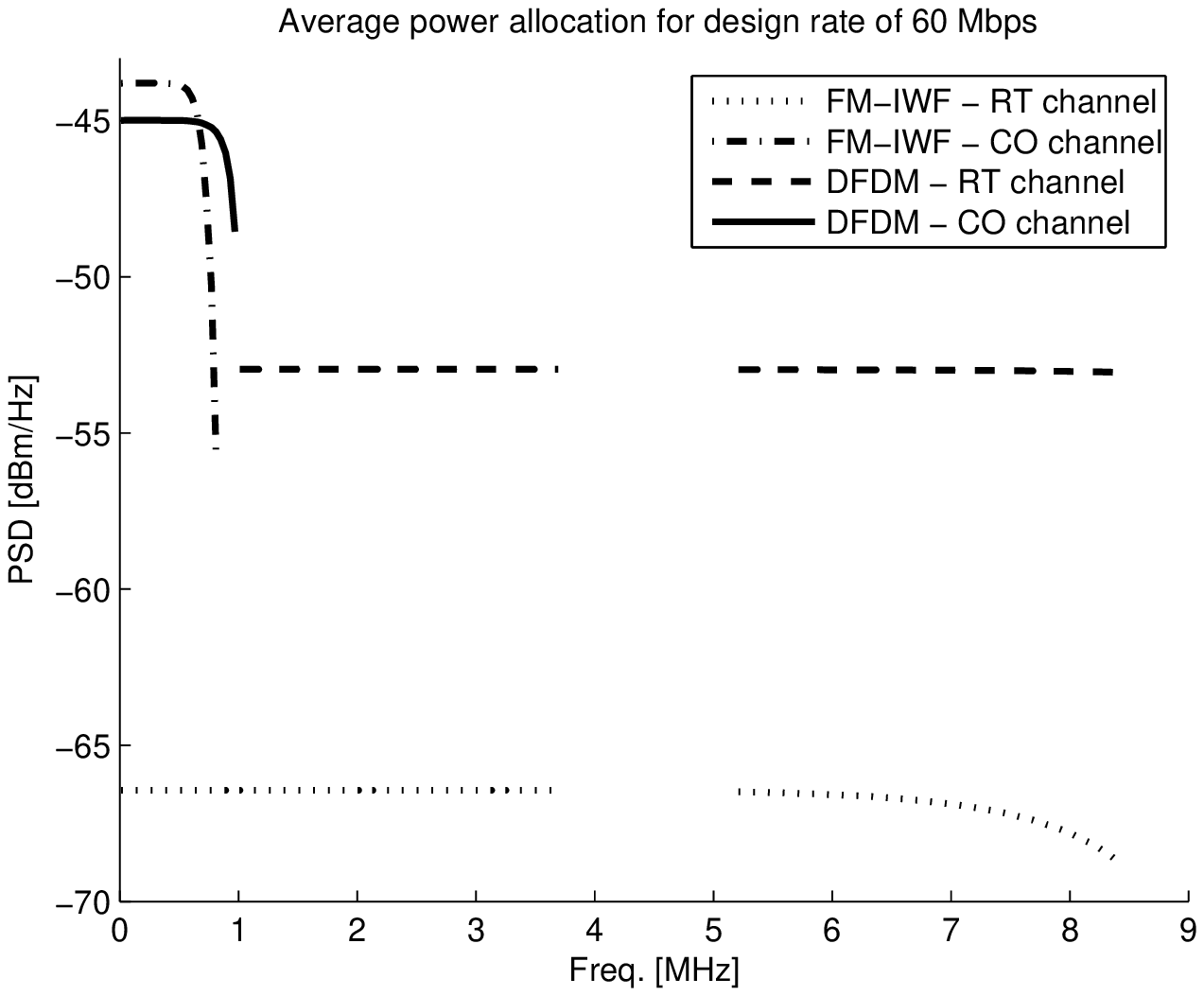}\label{downstream_psd}}
}
\caption{(\ref{downstream_rr})Downstream - Rate region for 8 VDSL
based RT and 8 ADSL based CO using FM-IWF and
DFDM. (\ref{downstream_psd}) Downstream PSD for FM-IWF and DFDM.}
\end{figure}

\begin{figure}[htbp]
\centerline
{
\subfigure[]{\includegraphics[width=6cm]{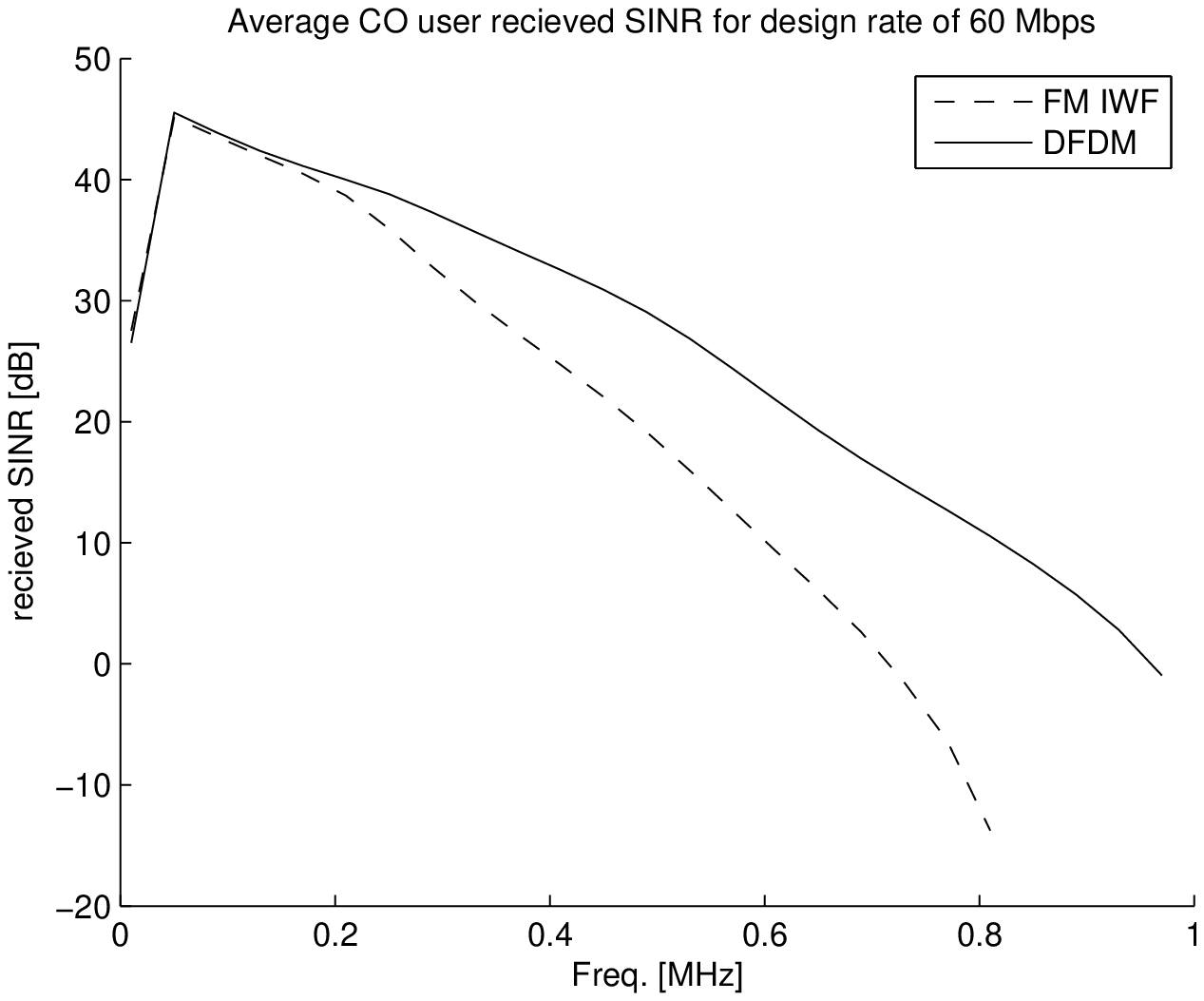}\label{downstream_sinr_a}}
\hfil
\subfigure[]{\includegraphics[width=6cm]{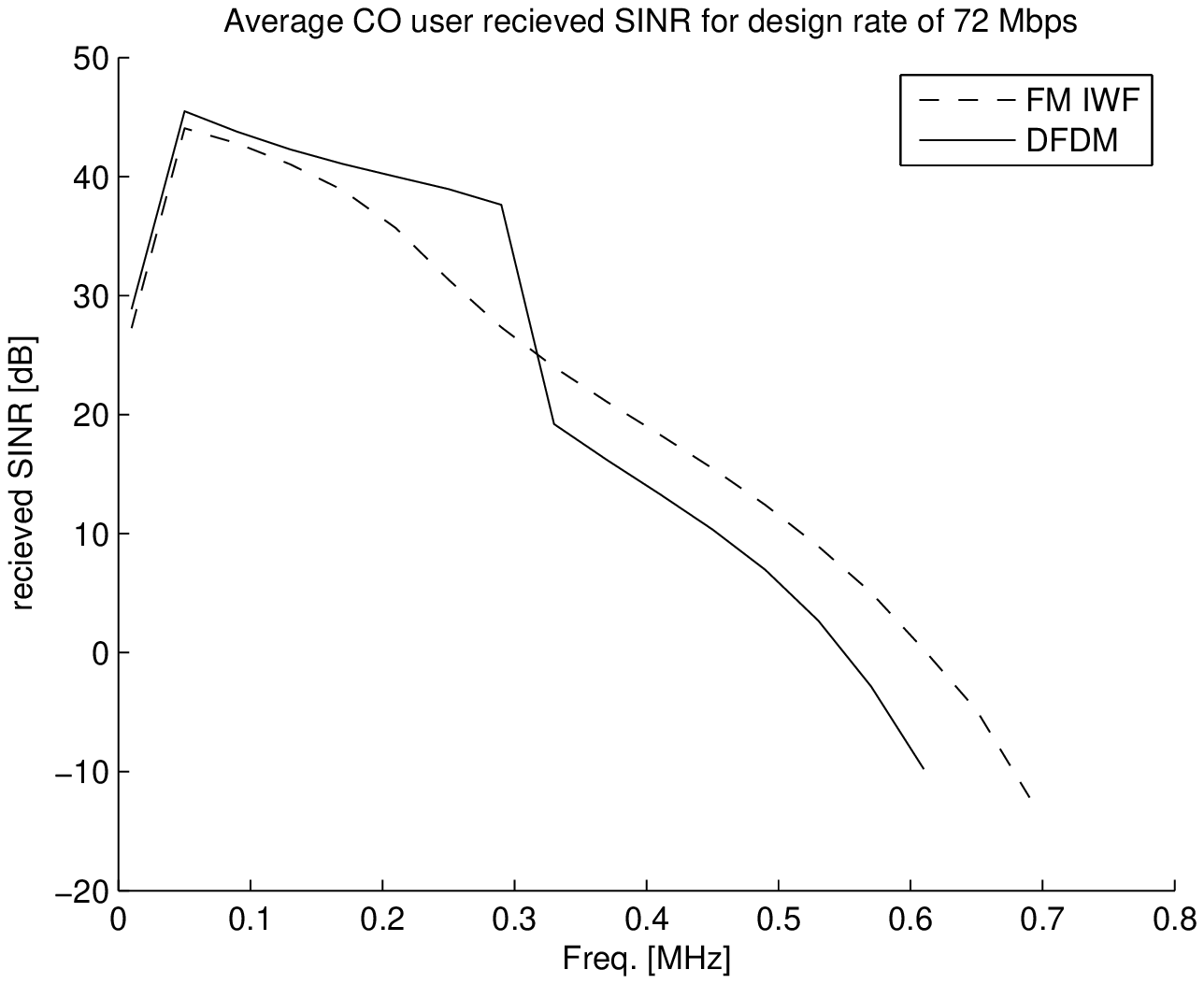}\label{downstream_sinr_b}}
}
\caption{ Average CO user received SINR for
FM-IWF and DFDM for RT user at 60 Mbps (\ref{downstream_sinr_a} ) and 72
Mbps (\ref{downstream_sinr_b} ) .}
\end{figure}

Looking at the DS scenario. The achieved rate regions of the three
methods are depicted in Figure \ref{downstream_rr}. We can clearly
see the advantage of the DFDM over the FM-IWF.  The PSDs of the DFDM
and the FM IWF methods corresponding to a 60 Mbps service on the RT
lines are shown on Figure \ref{downstream_psd}. For this value of
$R_d$ there is no overlap between the frequencies used by each
cluster of users resulting in no interference to CO users from RT
users. This is the best case for the CO users since actually the
near far problem has vanished and the achieved rate of the average
CO user is the same as the RT was not transmitting at all. Figure
\ref{downstream_sinr_a} shows the received SINR of an average CO
user for both methods. Its implies that for $R_d$ for which $f_c$ is
grater than the maximal frequency used by the CO users the gain
using DFDM has two factors. The first factor is that the DFDM's SINR
is grater or equal (since there is no interference from the RT) than
the FM IWF one. The second is that the CO users available bandwidth
is larger using DFDM than the FM IWF bandwidth. Both originate from
the orthogonality of the transmission bands and both factors have
positive contribute on the achieved rate of the CO users. Where
$R_d$ is close to the RT maximal achievable rate $f_c$ is getting
smaller and the available bandwidth for the CO is decreased. Figure
\ref{downstream_sinr_b} demonstrates this for $R_d=72$ Mbps. This
design rate is almost $0.93\cdot R_{RT,max}$ and thus even by
applying DFDM the RT PSD occupies most of the low frequencies
regime. This causes the bandwidth of the CO users to decrease to 0.6
Mhz and in addition to a degradation in the SINR. As a consequence
for this $R_d$ FM IWF achieves better rate for the CO users than
DFDM. However as can be seen the difference is marginal.

\begin{figure}[htbp]
\centering \epsfig{file=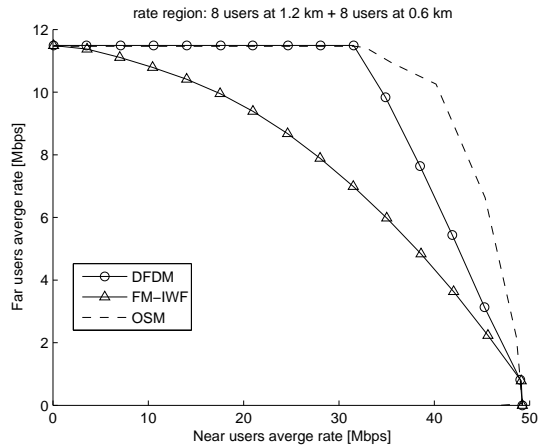,
width=0.45\textwidth} \caption{Upstream - Comparison of the rate
region for 8 Far users and 8 Near users using FM-IWF and DFDM.}
\label{upstream_rr}
\end{figure}

Turning to the upstream scenario. Figure \ref{upstream_rr} depicts the
rate region achieved by the different DSM methods. Not only the DFDM outperforms the FM IWF
method in this scenario, the rate region obtained by the DFDM method
is very close to the upper bound given by a fully coordinated spectrum
management using the OSM algorithm. Moreover in this scenario the DFDM
is better than or equal to the FM-IWF for all achievable rates of the
strong user.

\section{Conclusions}
In this paper we have analyzed the iterative water filling algorithm
for several simple channels using game theoretic techniques. We have
shown that the IWF algorithm is
subject to the prisoner's dilemma by providing explicit
characterization of its rate region for these cases. Based on these
insights we proposed a distributed coordination algorithm
improving the rate region in near-far scenarios. Finally we have
provided experimental analysis of these two algorithms and the optimal
centralized algorithm on measured channels.
\section*{Acknowledgement}
We would like to
thank Dr. Meryem Ouzzif and Dr. Rabah Tarafi  and Dr. Hubert Mariotte of France Telecom
R\&D, who conducted the VDSL channel measurements on behalf of France
Telecom R\&D under the auspices of the U-BROAD project.

\section{Appendix A: Proof of the existence of Nash equilibrium}
In this section we prove that for every sequence of intervals
$\{I_1,\ldots,I_k\}$ ,the Gaussian interference game has a Nash
equilibrium point. Our proof is based on the technique of
\cite{nikaido55}, (see also \cite{basar82}),  adapted to the
water-filling strategies in the
game GI. While the result follows from standard game theoretic results, it
is interesting to see the continuity of the water-filling
strategy as the reason for the existence of the Nash equilibrium.
\bthm
\label{nash}
For any finite partition $\{I_1,\ldots,I_k\}$ a Nash
equilibrium in the Gaussian interference game
$GI_{\{I_1,\ldots,I_k\}}$ exists.
\ethm
{\bf Proof:} For each player $i$ define the water-filling function
$W_i(\vp_1,\ldots,\vp_N): \mB \rarrow \mB_i$, which
is  the power distribution that maximizes $C^i$ given that for
every $j \neq i$ player $j$  uses the power distribution
$\vp_j$ subject to the power limitation $P_i$. The value of
$W_i(\vp_1,\ldots,\vp_n)$ is given by water-filling with total power
of $P_i$ against the noise power distribution composed of
\beq
N_i(k)=\frac{1}{|h_i(k)|^2}\left[\sum_{j \neq i} |h_ij(k)|^2 p_j(k)+n_i(k)\right]
\eeq
where for all $k$, $n_i(k)>0$ is the external noise power in the $k$'th band.
\bcl
$W_i(\vx_1,\ldots,\vx_N)$ is a continuous function.
\ecl
{\bf Proof:}
We shall not prove this in detail. However informally this fact is
very intuitive since small variations in the
noise and interference power distributions will lead to small changes in the
waterfilling response.
The proof of theorem \ref{nash} now easily follows from the Brauwer
fixed point theorem. The function $\mW=[W_1,\ldots,W_N]$ maps
$\mB$ into itself. Since $\mB$ is  compact subset of a finite
dimensional Euclidean space $\mW$ has a fixed point
$\left[\vp_1,\ldots,\vp_N \right]^T$. This means that
\[
\mW(\left[\vp_1,\ldots,\vp_N\right]^T)=\left[\vp_1,\ldots,\vp_N\right]^T
\]
By the definition of $\mW$ this means that each $\vp_i$ is the result
of player $i$ water-filling its power against the interference
generated by $\{\vp_j : j \neq i\}$
subject to its power constrain . Therefore
$\left[\vp_1,\ldots,\vp_N \right]^T$ is a Nash equilibrium for
$GI_{\{I_1,\ldots,I_K\}}$.

\section{Appendix B: Bounds on the rate region of dynamic FDM}
\label{appendixB}
We can now obtain similar equations defining a dynamic FDM strategy,
where the Bully uses the minimal fraction $\gl W_1$ of the first
band to achieve $R_2$. The main concept of this method is to
minimize the interference to the weak user. This translates to
minimize $\gl$ for any given $R_2$. As a consequence we will not
apply any power backoff (i.e. $\gt=1$) in order to maximize the
power at the second band. The minimization of $\gl$ is done through
the maximization of the achieved rate for any given $\gl$. Since the
noise PSD is equal for both bands (recall that we neglect the
interference from the weak user at the first band) maximizing the
rate is equal to waterfill the power along a new single band channel
with effective bandwidth of $\gl W_1+W_2$ where $\gl$ is chosen such
the following equation holds
\begin{equation}
\label{R2_fdm}
R_2 = (\gl W_1+W_2)\log_2\left(1+\frac{P}{(\gl
W_1+W_2)N_2}\right)
\end{equation}
In order to get upper and lower bounds on $R_1$ under the new
strategy we can bound the total used bandwidth

\beq W_2\leq \gl W_1+W_2 \leq W_1+W_2 \eeq
Thus we get
\begin{equation}
R_2 \leq (\gl W_1+W_2)\log_2\left(1+\frac{P}{W_2N_2}\right)
\end{equation}
and we derive  that $\gl \geq \gl_{\min}$, where
\begin{equation}
\label{L_lower}
\gl_{\min} =
\frac{\frac{R_2}{\log_2(1+\frac{SNR_2}{W_2})}-W_2}{W_1}
\end{equation}
on the other hand
\begin{equation}
R_2 \geq (\gl W_1+W_2)\log_2\left(1+\frac{P}{(W_1+W_2)N_2}\right)
\end{equation}
and similarly $\gl \leq \gl_{\max}$, where
\begin{equation}
\label{L_upper}
\gl_{\max} =
\frac{\frac{R_2}{\log_2(1+\frac{SNR_2}{(W_1+W_2)})}-W_2}{W_1}
\end{equation}
Recall that $0\leq \gl \leq 1$, if for given $R_2$ the obtained
$\gl$ is grater than 1 this implies that the given rate doesn't lie
in the achievable rate region of the bully. On the other hand a
negative $\gl$ implies that the bully can achieve the desired rate
by the use of the second band solely (i.e. we will set $\gl$ to
zero).

The rate of user I is achieved by water-filling in the first band.
This results in \beq \label{R1_rate} R_1 = (1-\gl) W_1 \log_2 \left(
1+\frac{\ga P_{1,1-\gl}}{(1-\gl)W_1 N_1}\right)+\gl W_1\log_2 \left(
1+\frac{\ga P_{1,\gl}}{\gl W_1 N_1+\gb P_1 }\right) \eeq In order to
evaluate this expression we first need to find $P_1$ which is user
II power allocation at the band $\gl W_1$. Moreover we need to
compute $\{P_{1,1-\gl},P_{1,\gl}\}$ the power allocation vector of
user I.
$P_1$ is the power allocation at the first band of a two bands
channel with equal noise PSD and with no power backoff. We have seen
above that in this case we get $P_1 = \tilde{\gr}P$ where
$\tilde{\gr}=\frac{\gl W_1}{\gl W_1+W_2}$ hence
\begin{equation} \label{P1_fdm}
P_1 =\frac{\gl W_1}{\gl W_1+W_2}P
\end{equation}
$P_{1,1-\gl}$ and $P_{1,\gl}$ are the power allocation of user I
along the two sub-bands at first band. Those parameters determined
by WF where we define the power level in each sub-band as
$\tilde{P}_{1,1-\gl}$ and $\tilde{P}_{1,\gl}$. Thus we have
\begin{equation}
\label{userI_wf}
\begin{array}{c}
\tilde{P}_{1,1-\gl} = \tilde{P}_{1,\gl} + \frac{P_1}{\gl W_1}\\
(1-\gl) W_1 \tilde{P}_{1,1-\gl} + \gl W_1 \tilde{P}_{1,\gl} = P
\end{array}
\end{equation}
The first equation in (\ref{userI_wf}) stands for the constant level
of power + noise at each sub-band while the second equation applies
the total power constraint. solving (\ref{userI_wf}) we get
\begin{equation}
\label{userI_wf1}
\begin{array}{l}
P_{1,1-\gl} = (1-\gl)W_1 \tilde{P}_{1,1-\gl} = P\left((1-\gl)W_1 +
\gl
(1-\gl) W_1^2 \frac{1}{\gl W_1+W_2}\right)\\
P_{1,\gl} = \gl W_1 \tilde{P}_{1,\gl} = P\left(\gl W_1 P - \gl
(1-\gl) W_1^2 \frac{1}{\gl W_1+W_2}\right)
\end{array}
\end{equation}
substituting (\ref{userI_wf1}) and (\ref{P1_fdm}) in (\ref{R1_rate})
we get
\beq \label{R1_rate2} R_1 = (1-\gl) W_1 \log_2 \left(
1+\frac{\ga P\left(1+\gl W_1 \frac{1}{\gl
W_1+W_2}\right)}{N_1}\right) +\gl W_1\log_2 \left( 1+\frac{\ga
P\left( 1-(1-\gl) W_1 \frac{1}{\gl W_1+W_2}\right)}{N_1+\gb
\frac{P}{\gl W_1+W_2}}\right)
\eeq
Since the first sub-band (i.e.
$(1-\gl)W_1$) is interference free it is clear that $R_1$ is
monotonically decreasing with $\gl$. Hence we can derive upper and
lower bounds on the achieved rate $R_1$ by substituting
(\ref{L_lower}) and (\ref{L_upper}) respectively.

\end{document}